\documentclass[a4paper]{article}

\usepackage{graphicx}      
\usepackage{amsmath,amssymb}
\begin{document}

\title{Stable compensators in parallel to stabilize arbitrary proper rational SISO plants}

\author{Abdul Hannan Faruqi\thanks{ahannanf20@iitk.ac.in} \and Anindya Chatterjee\thanks{anindya@iitk.ac.in, anindya100@gmail.com}}
\date{Mechanical Engineering, IIT Kanpur, 208016\\ \smallskip \today\\ \smallskip
{\em The paper has been submitted to an IFAC conference (decision awaited)}}

\maketitle              

\begin{abstract}
We consider stabilization of linear time-invariant (LTI) and single input single output (SISO) plants in the frequency domain from a fresh perspective. Compensators that are themselves stable are sometimes preferred because they make starting the system easier. Such starting remains easy if there is a stable compensator in parallel with the plant rather than in a feedback loop. In such an arrangement, we explain why it is possible to stabilize all plants whose transfer functions are proper rational functions of the Laplace variable $s$. In our proposed architecture we have (i) an optional compensator $C_s(s)$ in series with the plant $P(s)$, (ii) a necessary compensator $C_p(s)$ in parallel with $C_s(s)P(s)$, along with (iii) a feedback gain $K$ for the combined new plant $C_s(s)P(s)+C_p(s)$. We show that stabilization with stable $C_s(s)$ and $C_p(s)$ is {\em always} possible. In our proposed solution the closed-loop plant is biproper and has all its zeros in the left half plane, so there is a $K_0$ such that the plant is stable for $K>K_0$. We are not aware of prior use of parallel compensators with such a goal. Our proposed architecture works even for plants that are impossible to stabilize with stable compensators in the usual single-loop feedback architecture. Several examples are provided.
\end{abstract}

\newtheorem{defn}{Definition}
\newtheorem{rem}{Remark}
\newtheorem{exmp}{Example}

\section{Introduction}
In classical control of linear time-invariant (LTI), single input single output (SISO) systems with rational transfer functions, stabilization with a stable controller in a single feedback loop is possible only if the plant satisfies a parity interlacing property (PIP), as shown by \cite{youla1974single}.
\begin{defn}[PIP]
Between every pair of real non-negative zeros of the open loop plant, the number of real positive poles must be even.
\end{defn}

Such stabilization is called {\em strong}. Even if strong stabilization is possible in principle, it may be impractical if it requires compensators of very high order. For example, the plant
\begin{equation}
    \label{eq:diffpl}
    P(s)=\frac{s^2-2s+1.1}{(s+2)(s+3)(s-4)}
\end{equation}
technically satisfies the PIP. However, in the usual single feedback loop architecture, strong stabilization requires a compensator of very high order. We will return to this example below.

\begin{figure}[h]
    \centering
    \includegraphics[width = 0.6\linewidth]{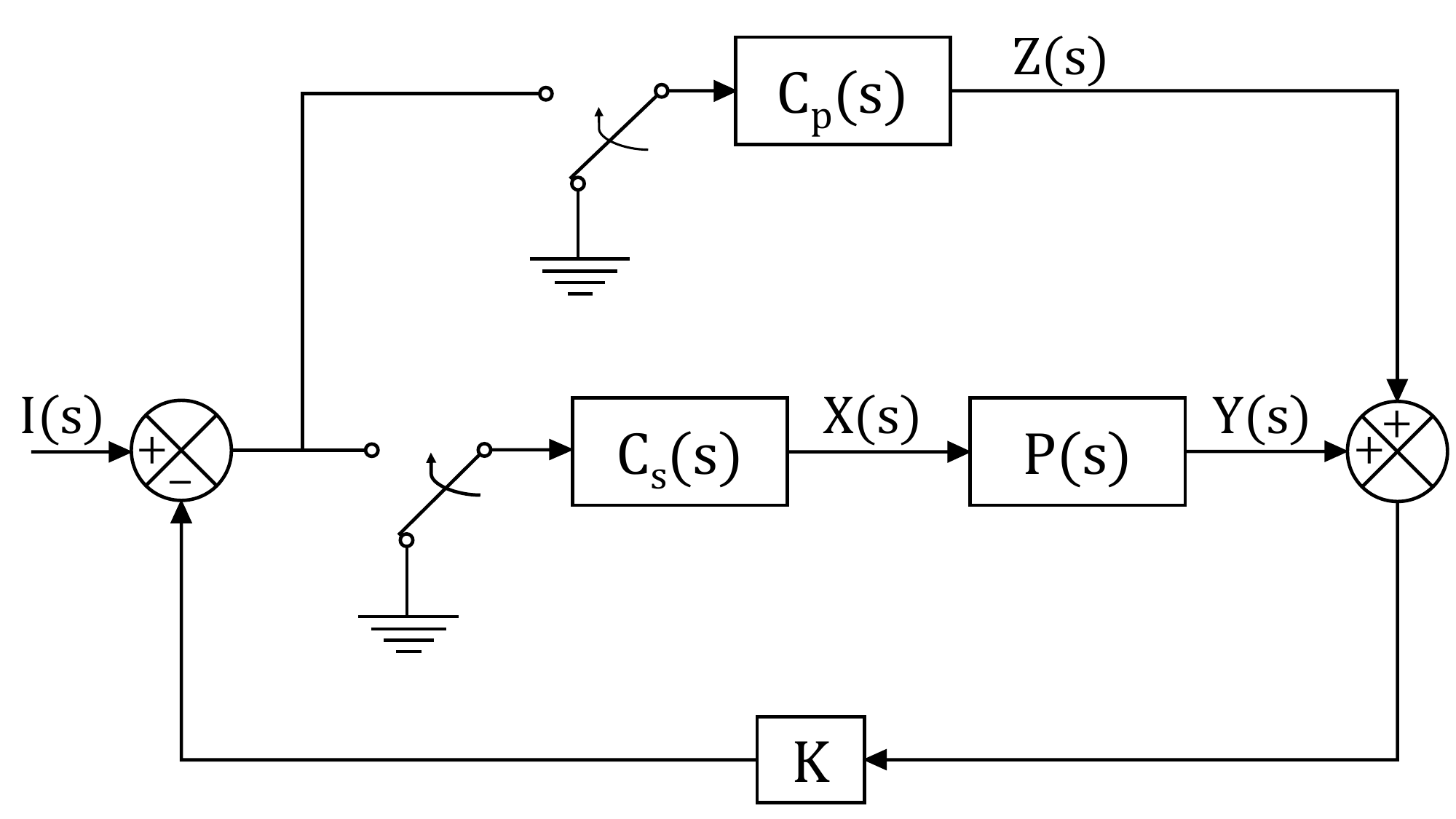}
    \caption{Our proposed architecture. $I(s)$ is an external input, $P(s)$ is the plant, $C_s(s)$ and $C_p(s)$ are stable compensators, and $K$ is a static gain. Two switches, operated simultaneously, can either engage the compensators or disengage them and ground their inputs.}
    \label{fig:BD}
\end{figure}

Our proposed control architecture is shown in
Fig.\ \ref{fig:BD}, along with two switches to indicate how starting the system can be done. The {\em stable} compensators $C_s(s)$ and $C_p(s)$ can be kept inactive with zero outputs until the plant $P(s)$ is brought close to its operating point, and then the compensators can be engaged by simultaneously closing the switches shown. In contrast, if a compensator is unstable, then setting its input to zero does not guarantee a bounded output, and {\em both} compensator and plant have to be brought to a suitable starting state before being released into coupled stable behavior. In this way, at least in principle, closed loop stability using stable $C_s(s)$ and $C_p(s)$ is advantageous for starting or restarting the system. However, to our knowledge, the architecture of Fig.\ \ref{fig:BD} has not been exploited from this viewpoint. 

We acknowledge that \cite{dudiki2018ff} and \cite{optff} have considered something superficially similar to our arrangement, but they assumed that $P(s)$ could be factored into the form $P_1(s)P_2(s)$, and that the outputs of $P_1(s)$ and $P_2(s)$ were both independently available, so that a compensator could be placed in parallel with $P_2(s)$. This is a strong assumption about the physical nature of the plant, and in any case it renders the plant {\em not} SISO. Furthermore, the parallel compensation is taken to be a static gain in both of these works, and in the latter there is further compensation with the goal of disturbance rejection. In contrast, in our proposed architecture, we assume the SISO plant cannot be separated into a product of two transfer functions, our parallel compensation involves a transfer function computed using ideas of strong stabilization, and we have concentrated on stabilization alone.

In this work, we show that for SISO plants whose transfer functions are proper rational functions of the Laplace variable $s$, regardless of whether they satisfy the PIP or not, stabilization is possible with stable $C_s(s)$ and $C_p(s)$ without pole-zero cancellation in the right half plane.

It may be noted that although the individual compensators $C_s(s)$ and $C_p(s)$ are stable, there is an ``effective'' compensator that may well be unstable.

\section{The effective compensator}
Referring to Fig.\ \ref{fig:BD}, assuming the switches are closed and there is no external input, we can find the effective compensator as follows. Let the output of $C_p(s)$ be $Z(s)$ and let the input to 
$P(s)$ be $X(s)$. Then
\begin{align*}
    Z&=-KC_p(Y+Z), \mbox{ and}\\
    X&=-KC_s(Y+Z),
\end{align*}
whence
$$X = \frac{-KC_s}{1+KC_p} Y.$$
In contrast, if we just had some single compensator $\Bar C$ in a single feedback loop, we would have
$X = - \Bar C Y$. Thus, our effective compensator is
$$C_{\rm eff} = \frac{KC_s}{1+KC_p},$$
which may be unstable. However, as indicated by the switches in Fig.\ \ref{fig:BD}, the individual stable compensators can be kept disengaged until the plant is ready, which can make starting the full system easier.

\section{Finding the compensators}
We observe that whether a given transfer function is proper or improper is independent of whether it is stable or unstable. For example,
$$\frac{s-1}{s^2+s+4}$$
is stable and proper, while its inverse,
$$\frac{s^2+s+4}{s-1},$$
is unstable and improper. Improper transfer functions are unphysical, but they will be useful in our mathematical calculations below. Unstable transfer functions have poles in the right half plane. Our compensators, and our final controlled closed-loop plant, must be both proper and stable.

Now consider a plant $P(s)$ with the co-prime factorization
$$P(s) = \frac{N(s)}{D(s)},$$
where both $N(s)$ and $D(s)$ are proper and stable, as usual.

Let us first ignore $C_s(s)$ and observe the consequences of $C_p(s)$, which is
a stable compensator in parallel. The combined transfer function is
\begin{align*}
    R(s) &= \frac{N(s)}{D(s)} + C_p(s)\\
    &= \frac{N(s) + D(s)\,C_p(s)}{D(s)}
\end{align*}
Multiplying by $D(s)$ gives
\begin{equation}
    U_p(s) = N(s) + D(s)\,C_p(s)
\end{equation}
The above expression is similar to the one used for the interpolating transfer function $U(s)$ for strong stabilization in the standard approach with a compensator in a single feedback loop (see e.g., \cite{doyle2013feedback,faruqi2023strong}). The only difference is that $D(s)$ and $N(s)$ are interchanged.

Mathematically speaking, if we were stabilizing the (possibly improper) plant
\begin{equation} \label{gg11} \frac{1}{P(s)} = \frac{D(s)}{N(s)}
\end{equation}
and found a suitable interpolant $U_p(s)$,
and if we then used the compensator 
\begin{equation}
    \label{eq:Cp}
    C_p(s) = \frac{U_p(s)-N(s)}{D(s)}
\end{equation}
in a single feedback loop, then the plant of Eq.\ (\ref{gg11}) would be stabilized, i.e., all of its poles would be in the left half plane. It follows that if we placed the same compensator in {\em parallel} with the original plant $P(s)$, then all the {\em zeros} of the combined plant $C_p(s)+P(s)$ would be in the left half plane. If, in addition, the combined plant $C_p(s)+P(s)$ was biproper, then it would have as many poles as it had zeros, and then a suitably large feedback static gain $K$ would {\em guarantee} stability from elementary ideas of root locus plots.

The above equivalence is the key insight that drives our present paper. We summarize its implications for clarity below:
\begin{rem}
    Stabilizing the inverse of a plant, say $1/P(s)$, with a compensator $C(s)$ in a single feedback loop is mathematically identical to using the same compensator in parallel with the plant so that the zeros of $C(s)+P(s)$ are all in the left half plane. In particular, methods for the former, which moves poles to the left half plane, can be used for the latter, which moves zeros to the left half plane.
\end{rem}

We can thus compute a suitable $U_p(s)$ from a {\em feedback} point of view, a solution for which is guaranteed to exist for plants that satisfy the PIP. In our case that means the inverse of the plant, i.e.,
Eq.\ (\ref{gg11}), should satisfy PIP: we will say more about this soon. For computing $U_p(s)$, several approaches have been proposed in the literature (see e.g., \cite{faruqi2023strong, Hitayhinf} and references therein). Of these, a method called AFP, or adjustment of fractional powers, is presented for plants of relative degree less than 3 in \cite{faruqi2023strong}. Here, since $P(s)$ is assumed proper, the plant in question (Eq.\ (\ref{gg11})) has relative degree less than or equal to zero, and AFP is guaranteed to work. Any other method of strong stabilization can be used in its place if the user so wishes, although the examples below have primarily used AFP.

We now come to the role of $C_s(s)$. For a viable $U_p(s)$ to exist, as noted above, the inverse of the plant $P(s)$ must satisfy the PIP. In other words, the plant must satisfy an inverse parity interlacing property (IPIP):
\begin{defn}[IPIP]
    Between every pair of real non-negative poles of the open loop plant, the number of real positive zeros must be even. Note that proper plants do not have poles at infinity.
\end{defn}

\begin{rem}
    As mentioned above, and as will be seen from the examples below, the fact that $1/P(s)$ is improper will have no consequence because it will be used for intermediate calculations alone. 
\end{rem}

\begin{rem}
    Since $U_p(s)$ and $D(s)$ are always biproper (see \cite{faruqi2023strong} for discussion), the compensator obtained from Eq.\ \ref{eq:Cp} will also be proper, irrespective of the relative degree of $N(s)$. 
\end{rem}

Now the role of $C_s(s)$ is clear. For a plant that does not satisfy the PIP, it is not possible to add right half plane {\em poles} using a stable premultiplying transfer function. However, for a plant that does not satisfy the IPIP, we can always use a stable premultiplying transfer function to add {\em zeros} in the right half plane. In this way, stable $C_s(s)$ can always be used to make any rational plant $P(s)$ satisfy the IPIP.
In our examples below, $C_s(s)$ is taken to be biproper and used only when necessary.

We now present several examples to show that our proposed stabilization method works easily for a variety of difficult plants. We begin with some hypothetical plants, but end with plants corresponding to specific mechanical systems.

\section{Examples}
\begin{exmp}\label{ex1} Plant that violates both PIP and IPIP.\end{exmp}

Consider the plant
\begin{equation}
    P(s) = \frac{(s-1)(s-3)}{(s-2)(s-4)}.
\end{equation}
This plant has a right half plane pole at  $s=2$, which is between zeros at $s=1$ and $s=3$. Therefore, it does not satisfy the PIP, and stabilization with a stable compensator in a single feedback loop is {\em not} possible. The plant does not satisfy the IPIP either, because there is a zero at $s=3$ lying between the poles at $s=2$ and $s=4$.
However, choosing the {\em stable}
\begin{equation}
    C_s(s) = \frac{s-2.8}{s+5},
\end{equation}
the modified plant
\begin{equation}
    \Bar{P}(s) = C_s(s)P(s) = \frac{(s-1)(s-2.8)(s-3)}{(s+5)(s-2)(s-4)}
\end{equation}
satisfies the IPIP.
Using the AFP method of \cite{faruqi2023strong} for $1/{\Bar{P}(s)}$, we find a $3^{\rm rd}$ order stable compensator,
$$C_p(s) = \frac{-134.09 (s+45.38) (s-1.103)}{(s+80.38) (s+78.77) (s+5)}.$$
Placing this in parallel with $\Bar{P}(s)$, we obtain a $5^{\rm th}$ order biproper plant (not given here for reasons of space) which has 5 poles and also 5 zeros. The poles are at
$$-80.38,
  -78.77,
   -5,
    2, \mbox{ and } 4.$$ Thus, there are 2 right half plane poles.
The zeros are at
$$-5.8775 \pm 1.6335i,
  -2.5999 \pm 1.9296i, \mbox{ and }
  -1.3053,$$
which are all in the left half plane.
Thus, as expected,
with $K$ in the feedback loop, the closed loop plant is stable for $K > 380$.

\begin{exmp}\label{ex2} Difficult to stabilize plant that satisfies both PIP and IPIP.\end{exmp}

We consider the plant given earlier in Eq. \eqref{eq:diffpl},
\begin{equation}
    P(s) = \frac{s^2 - 2 s + 1.1}{(s+2) (s+3)(s-4)}.
\end{equation}
This plant has complex conjugate zeros close to the positive real axis. If, under small perturbation, those zeros became a real pair, then the plant would violate PIP. It is therefore a difficult plant to control in the usual single feedback loop configuration (see \cite{SMITH1986127, Hitayhinf}).
However, the plant satisfies the IPIP, and is easy to stabilize using our approach.
Note that the inverse is not proper; as emphasized earlier, this makes no difference to the stabilization calculation. Since the plant satisfies IPIP, $C_s(s)$ is not needed and is taken as 1. Using the AFP method to stabilize $1/P(s)$, we find a $3^{\rm rd}$ order stable $C_p(s)$ as given in Appendix \ref{res}. Placing this compensator in parallel with the plant as per our theory, the combined plant $C_p(s)+P(s)$ is $4^{\rm th}$ order and biproper in this case. There is one right half plane pole, but all 4 zeros are in the left half plane. With feedback, the system is stable for $K>180$. 
We mention that direct feedback stabilization with the usual single feedback loop is in principle possible for this plant, but the compensator has high order. With the AFP method, we found a stabilizing stable feedback compensator of order 39: this number is representative although possibly not minimal. Other methods (such as Nevanlinna-Pick interpolation as in \cite{Hitayhinf}) yield very high orders also. Details are omitted.

\begin{exmp}\label{ex3} Another plant that satisfies PIP and is difficult to stabilize. \end{exmp}

We consider a plant which satisfies the PIP, but for which the AFP method applied to the usual single loop feedback configuration yields a compensator of order 36 (see \cite{faruqi2023strong}). The plant is
\begin{equation}
    P(s) = \frac{ (s^2 - 4s + 40)^2}{(s+2) (s-4) (s+6) (s+8) (s+10) (s+12)}.
\end{equation}
This plant is difficult to stabilize with the usual single loop feedback structure in the sense that calculations are slightly involved because it has 2 repeated right half plane zeros.

Since the plant satisfies IPIP, we choose $C_s(s) = 1$. Then, using the AFP method for $1/P(s)$ the compensator obtained is $7^{\rm th}$ order and is given in Appendix \ref{res}. Placing that compensator in parallel with $P(s)$ and with a feedback loop as before, the closed loop system is stable for $K>2.36 \times 10^4$. Further details are omitted.

\begin{exmp}\label{ex5} A plant with symmetrically placed poles and zeros.\end{exmp}

Consider the following plant having 4 right half plane zeros, with 4 left half plane poles placed symmetrically, along with a single right half plane pole:
\begin{equation}
    P(s) = \frac{(s^2 - 2s + 2) (s^2 - 2s + 5)}{(s-3) (s^2 + 2s + 2) (s^2 + 2s + 5)}.
\end{equation}
The plant satisfies IPIP, so we choose $C_s(s) = 1$. Then, using the AFP method for $1/P(s)$, we get a $5^{\rm th}$ order compensator (see Appendix \ref{res}). Placing that compensator in parallel with $P(s)$ as per our theory,
with feedback we obtain stability for $K>196$.
In the classical single-loop feedback framework, the Nevanlinna-Pick interpolation-based method of \cite{Hitayhinf} gives a stabilizing stable compensator of order 56 for this plant, while our AFP method gives an even higher order.
Details are omitted.

\begin{exmp}\label{ex6}Inverted pendulums on a cart.\end{exmp}

\begin{figure}[h]
    \centering
   \includegraphics[width = 0.5\linewidth]{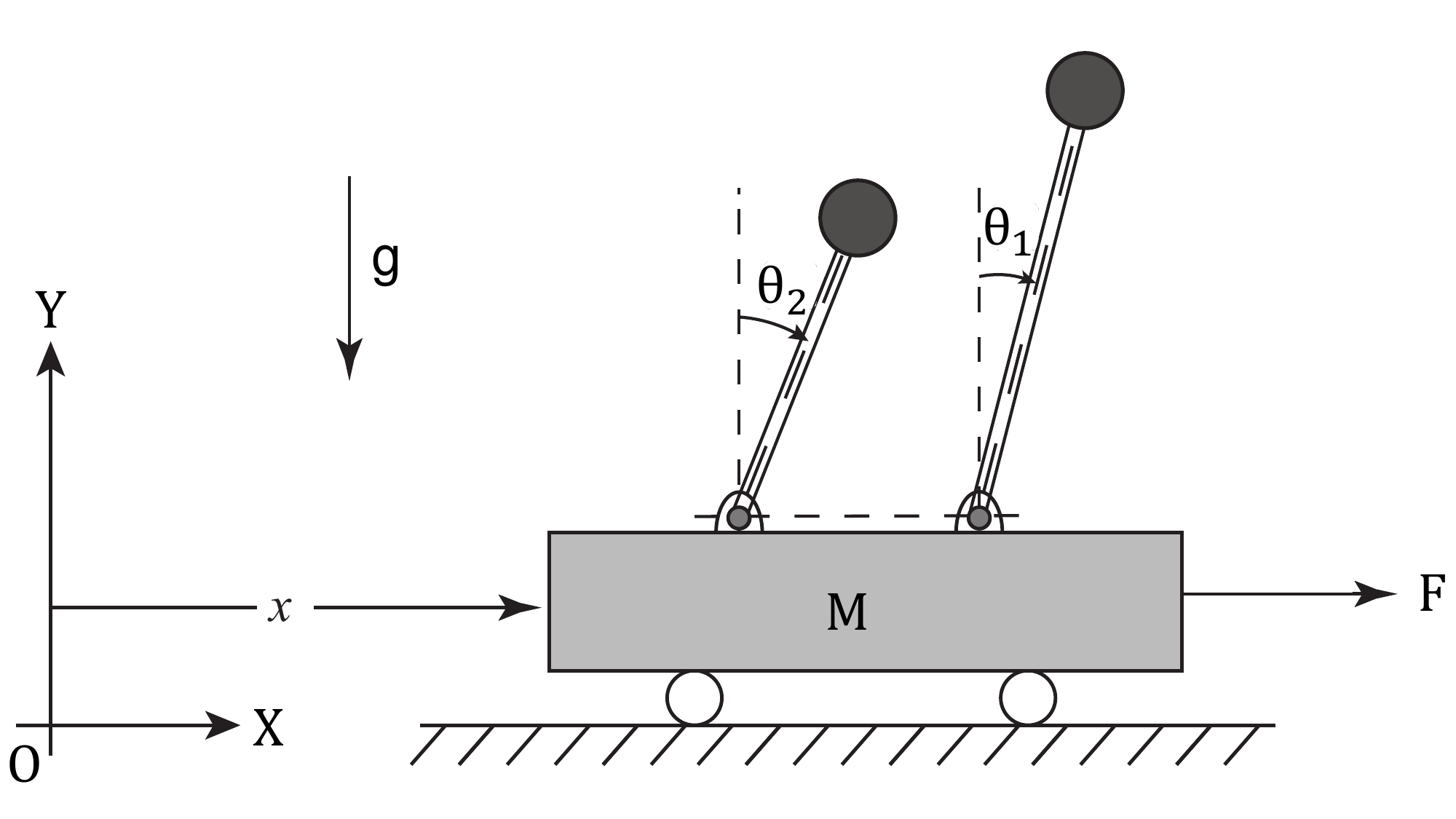}
    \caption{Two inverted pendulums on a cart}
    \label{fig:ipc}
\end{figure}

We now consider a mechanical system. Figure \ref{fig:ipc} shows two inverted pendulums on a cart, where the control input is a force on the cart.

We will approach this problem in two steps. First, for simplicity, let us begin with the simpler case of a cart with a single inverted pendulum. This reduced problem, for some parameter values, was
discussed by \cite{goswami2023balancing} and we consider the same parameters here.
The measured variable is {\em not} pendulum angle but cart position, which makes the problem more challenging. The resulting transfer function is
\begin{equation}
    P(s) = \frac{s^2-1}{s^2(0.3s^2-1.3)},
\end{equation}
which does not satisfy PIP. In \cite{goswami2023balancing}, the system was stabilized with stable compensators in series and parallel, obtained {\em via} a genetic algorithm based optimization calculation. We use our direct, although not optimal, approach for this problem.
We note that the plant does not satisfy IPIP. So, we use
\begin{equation}
    C_s(s) = \frac{s-1.5}{s+15}
\end{equation}
where the choice of the ``15'' in the denominator is arbitrary, and which gives
\begin{equation}
    \Bar{P}(s) = \frac{(s-1.5)(s^2-1)}{s^2(0.3s^2-1.3)(s+15)}
\end{equation}
Now applying AFP, we get a $11^{\rm th}$ order compensator for $1/\Bar{P}(s)$ (see Appendix \ref{res}). Putting that compensator in parallel with $\Bar{P}(s)$ as per our theory, we move all zeros to the left half plane. Subsequently, feedback with $K>1.25 \times 10^4$ stabilizes the combined system.

We now consider the problem with two inverted pendulums, but with the pendulum angle $\theta_1$ being measured. The tradeoff is that adding a pendulum makes the system harder to stabilize, but supplying a pendulum angle measurement rather than cart position measurement makes the system easier to stabilize.

In terms of mechanical parameters, the transfer function for this system is
\begin{equation}
    P(s) = {(-g+s^2L_2)\over\displaystyle{L_2L_1M\,s^4 -g(L_1M+L_1m_2+L_2M+L_2m_1)s^2\atop \quad {} + g^2(M+m_1+m_2)}}
\end{equation}
where $m_1,m_2$ are pendulum masses,  $L_1,L_2$ are pendulum lengths, and $M$ is the mass of the cart. Assigning parameter values $g = 9.8, M = 2, m_1 = 0.5, m_2 = 0.5, L_1 = 1.2$, and $L_2 = 0.8$, we obtain
\begin{align}
    P(s) 
    &= \frac{ -0.41667 (s-3.5) (s+3.5)}{(s+4.041) (s+3.031) (s-3.031) (s-4.041)}.
\end{align}
The system violates PIP and cannot be stabilized with a stable compensator in the classical framework. It violates IPIP also, with zero at 3.5 between poles at 3.031 and 4.041. We use 
\begin{equation}
\label{zxc}
    C_s(s) = -2.4\left(\frac{s-3.5}{s+3.5}\right),
\end{equation}
which cancels a left half plane zero (allowed).
The modified plant becomes
\begin{equation}
    \Bar{P}(s) = \frac{(s-3.5)^2 }{(s^2-16.3333)(s^2-9.1875)},
\end{equation}
where the numerator and denominator polynomials of $\Bar{P}(s)$ are rendered monic due to the inclusion of ``$-2.4$" in Eq.\ (\ref{zxc}).
$\Bar{P}(s)$ satisfies IPIP and the AFP method applied to $1/\Bar{P}$ gives a $10^{\rm th}$ order compensator (see Appendix \ref{res}). Placing that compensator in parallel with $\Bar{P}(s)$ as per our theory moves all zeros of the resulting plant to the left half plane. Subsequently, stability is achieved using a feedback gain $K>6.8\times 10^4$.

\section{Conclusions}
While there is no unequivocally superior way to stabilize a general unstable system using feedback control, in this paper we have presented and demonstrated a way of doing so that may be preferable for some systems. In the specific case of LTI SISO systems whose transfer functions are proper rational functions of the Laplace variable $s$, we have shown that a {\em general} method of stabilization is to use an optional compensator $C_s(s)$ in series with the plant $P(s)$, and then to use a feedforward compensator $C_p(s)$ in parallel with $C_s(s)P(s)$, such that the combined plant $C_s(s)P(s)+C_p(s)$ is biproper and also has all its zeros in the left half plane. Stabilization of the combined plant is then easy: simple feedback with a large enough static gain $K$ guarantees stability.

The motivation for our proposal is that a control architecture with stable compensators, possibly in both series and parallel, allows simpler starting and restarting of the plant. Furthermore, some plants cannot be stabilized with stable compensators in the usual single feedback loop architecture, but they can be stabilized in our proposed architecture with stable compensators. An ``effective'' compensator computed for our prposed arrangement may well be unstable even for plants that satisfy the PIP, but as explained, the individual compensators are stable and can be kept turned on yet disconnected until the user is ready to engage them.

The insight that drives our proposed approach is that designing a parallel compensator that will move zeros of a plant to the left half plane is the same as designing a usual feedback compensator that will move the poles of the {\em inverse} of the same plant to the left half plane.  The advantage of our approach lies in that adding right half plane zeros to a plant is possible with a stable series compensator $C_s(s)$, and such zeros can be used to make the plant satisfy the inverse parity interlacing property (IPIP). An attempt to introduce right half plane poles with a series compensator would necessitate an {\em unstable} compensator, which we avoid.

In our approach, the first part is easy. In the first part, to decide whether or not the series compensator $C_s(s)$ is necessary, we just check if the plant satisfies IPIP. Further, designing $C_s(s)$ is easy as well, because we just need to add right half plane zeros within inconvenient intervals wherein odd numbers of such zeros appear. The only significant computation in our approach is that of $C_p(s)$, which can be attempted in any way preferred by the user. In this paper, for computing $C_p(s)$, we have primarily used a method called adjustment of fractional powers (AFP), which is guaranteed to work, and which we have presented elsewhere.

\section*{Acknowledgement}
    Kundan Maji helped with Fig.\ \ref{fig:ipc}. Dennis Bernstein explained how the effective compensator
    can be unstable even though individual parts are not.

\bibliographystyle{ieeetr}
\bibliography{citations}

\appendix
\section{Compensators from the examples}
\label{res}
\begin{enumerate}
    \item Example \ref{ex2}
    $$C_p(s) = \frac{(s+18.17) (s-4.657) (s-0.5159)}{(s+180.6) (s+3) (s+2)}$$
    \item Example \ref{ex3}
    $$C_p(s) = \frac{(s^2 - 1.957s + 18.18) (s^2 + 2.56s + 59.16)(s+22.45)(s^2 + 24.95s + 315.1)}{(s+138)^2 (s+2)(s+6) (s+8)(s+10) (s+12)}$$
    \item Example \ref{ex5}
    $$C_p(s) = \frac{(s+18.72) (s-4.584) (s-2.401)(s^2 - 1.739s + 3.202)}{(s+201) (s^2 + 2s + 2)(s^2 + 2s + 5)}$$
    \item Example \ref{ex6}\\
    Part 1
    \begin{align*}
    C_p(s) &= \frac{(s+1.39) (s+1) (s-30.99)(s^2 + 2.425s + 1.594)}{(s+15)(s+271.1)^3 (s+2.082)}\\
    & \times \frac{(s^2 + 86.56s + 1944) (s^2 + 69.75s + 2188)(s^2 - 3.429s + 1773)}{(s+1.087)^4(s+59.87)^2}
    \end{align*}

    \noindent Part 2
    \begin{align*}
        C_p(s) &= \frac{(s^2 + 2.139s + 1.153) (s^2 + 130.2s + 4599)(s^2 + 78.13s + 4132) (s^2 - 21.6s + 2924)}{(s+3.031) (s+1)^3(s+297.8)^4 }\\
        & \times \frac{(s-41.73)(s+0.8532)}{(s+4.041)(s+297.8)}
    \end{align*}
\end{enumerate}

\end{document}